\documentclass{aastex63} 

\usepackage[capbesideposition=right]{floatrow}
\usepackage{wrapfig}
\usepackage[right = 0.5in, bottom = 1in]{geometry}

\newcommand\aastex{AAS\TeX}

\newcommand{\Rs}{$R_{\odot} $}
\def\ion[#1 #2]{#1\,{\sc #2}}

\submitjournal{ApJ}
\shorttitle{\aastex\ Absolute Brightness and MHD Model Predictions of Fe X, XI and XIV}
\shortauthors{Boe et al.}

\begin{document}

\title{The Solar Minimum Eclipse of 2019 July 2: II. The First Absolute Brightness Measurements and MHD Model Predictions of Fe X, XI and XIV out to 3.4 \Rs}

\author{Benjamin Boe}
\affil{ Institute for Astronomy, University of Hawaii, Honolulu, HI 96822, USA}

\author{Shadia Habbal}
\affil{ Institute for Astronomy, University of Hawaii, Honolulu, HI 96822, USA}

\author{Cooper Downs}
\affil{Predictive Science Inc., San Diego, CA, 92121, USA}

\author{Miloslav Druckm\"uller}
\affil{Faculty of Mechanical Engineering, Brno University of Technology, Technicka 2, 616 69 Brno, Czech Republic}

\correspondingauthor{Benjamin Boe}
\email{bboe@hawaii.edu}
\begin{abstract}
We present the spatially resolved absolute brightness of the \ion[Fe x], \ion[Fe xi] and \ion[Fe xiv] visible coronal emission lines from 1.08 to 3.4 \Rs, observed during the 2019 July 2 total solar eclipse (TSE). The morphology of the corona was typical of solar minimum, with a dipole field dominance showcased by large polar coronal holes and a broad equatorial streamer belt. The \ion[Fe xi] line is found to be the brightest, followed by \ion[Fe x] and \ion[Fe xiv] (in disk $B_\odot$ units). All lines had brightness variations between streamers and coronal holes, where \ion[Fe xiv] exhibited the largest variation. However, \ion[Fe x] remained surprisingly uniform with latitude. The Fe line brightnesses are used to infer the relative ionic abundances and line of sight averaged electron temperature ($T_e$) throughout the corona, yielding values from 1.25 -- 1.4 MK in coronal holes up to 1.65 MK in the core of streamers. The line brightnesses and inferred $T_e$ values are then quantitatively compared to the PSI Magnetohydrodynamic model prediction for this TSE. The MHD model predicted the Fe lines rather well in general, while the forward modeled line ratios slightly underestimated the observationally inferred $T_e$ within 5 to 10 $\%$ averaged over the entire corona.  Larger discrepancies in the polar coronal holes may point to insufficient heating and/or other limitations in the approach. These comparisons highlight the importance of TSE observations for constraining models of the corona and solar wind formation.
\end{abstract} 


\section{Introduction} 
\label{intro}
Coronal emission lines were first discovered and identified in the visible and near infrared during total solar eclipses (TSEs) in the late 19th \citep{Young1872} and early 20th centuries \citep{Lyot1939}. Subsequent observations focused on the so-called ``Green" (\ion[Fe xiv], 530.3 nm) and ``Red" (\ion[Fe x], 637.4 nm) lines (e.g. \citealt{MagnantCrifo1973,Chandrasekhar1984, Bessey1984, Guhathakurta1992}). The observed emission was typically limited to a helioprojective distance less than 1.4 \Rs, though occasionally emission was recorded up to 1.7 \Rs \ \citep{Singh1982}. Recent work has continued to employ line emission observed during TSEs as a means to study the physical properties of the solar corona, including the ionic freeze-in distances \citep{Habbal2007, Habbal2013, Boe2018}, the average electron temperature ($T_e$; \citealt{Boe2020a}), the sources of the solar wind in the corona \citep{Habbal2021}, and the presence of various ionic species with slit-spectrographs \citep{Ding2017, Samra2018, Koutchmy2019}.

\par
Given the diagnostic potential of coronal emission lines, and the relative sparsity of TSEs, coronagraphs have been often utilized to study coronal line emission. \cite{Lyot1932} was the first to use a coronagraph to simulate a TSE, and was followed by systematic observations of \ion[Fe xiv] in the very low corona ($\approx$ 1.15 \Rs) over several decades \citep{Rybansky1994, Altrock2011}. More recent observations have utilized the near-infrared \ion[Fe xiii] (1074.4 nm) line (e.g. \citealt{Dima2019,Ruminska2022}), especially to measure the coronal magnetic field. In the near future, additional observations of numerous coronal visible lines will be performed with the ground-based UCoMP instrument \citep{Tomczyk2021}. 
\par
Space-based coronagraphs have also been utilized to study visible lines to some extent, including measurements of the line brightness \citep{Wang1997,Srivastava2000} and line-widths \citep{Mierla2008} of both \ion[Fe xiv] and \ion[Fe x] with the LASCO-C1 instrument in the low corona ($\lessapprox 1.5$ \Rs). Unfortunately, these observations were not absolutely calibrated and the data from C1 were limited due to its failure early in the SOHO mission. Notwithstanding, the observations of visible wavelength coronal lines have been rather limited outside the prevue of TSEs and are especially limited in spatial extent to 1.7 \Rs \ at most. (See \cite{DelZanna2018a} for a detailed historical overview of visible and near-infrared coronal line emission observations).
\par

In this work, we analyze observations of the \ion[Fe x], \ion[Fe xi] and \ion[Fe xiv] emission lines from the 2019 July 2 TSE. The observation and calibration procedures are discussed in Section \ref{Eclipse}. In Section \ref{Line} we discuss the spatially resolved absolute brightness of each emission line throughout the corona from 1.08 to 3.4 \Rs. In Section \ref{Temp} we infer the spatially resolved Line-of-sight (LOS) averaged electron temperature $T_e$ via the Fe line ratios. Finally, in Section \ref{PSI} we compare the line emission and $T_e$ values to the predictions of the Predictive Science Inc. (PSI) Magnetohydrodynamic (MHD) simulation of this eclipse. A discussion and summary of the results is given in Section \ref{conc}.

\section{2019 Total Solar Eclipse Observations}
\label{Eclipse}

\begin{figure*}[t!]
\centering
\includegraphics[width = 7in]{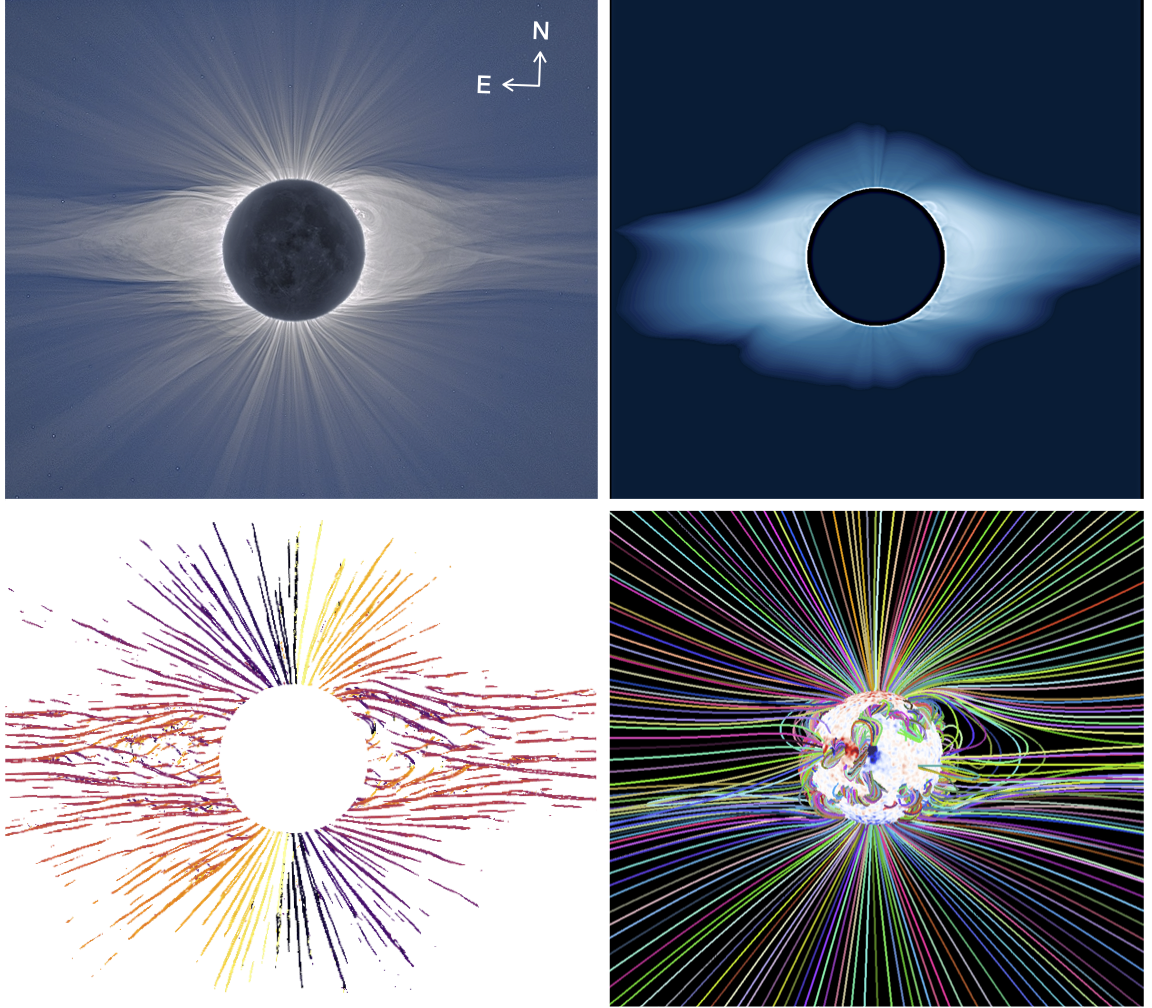}
\caption{Top Left: High spatial resolution white-light image of the eclipse corona, with terrestrial north pointing upward. The compass indicates solar north which is two degrees clockwise of terrestrial north. Top Right: PSI MHD model prediction of the white-light corona (see Section \ref{PSI}), with the same orientation and scale as the eclipse image to the left. Bottom left: Quantitative trace of the eclipse magnetic field morphology, made with the white-light image, previously presented in \cite{Boe2020b}. Bottom right: Traces of the coronal magnetic field lines from the PSI MHD model prediction of the eclipse, aligned to the white light image.}
\label{fig1} 
\end{figure*}

The eclipse observations used in this work were acquired during the 2019 July 2 Total Solar Eclipse over Rodeo, Argentina, where totality lasted 2 minutes and 14 seconds. The high spatial resolution broadband white light image at the top left panel of Figure \ref{fig1} highlights the magnetic morphology of solar minimum with a dipolar field dominance. Specifically, the corona consisted of large polar coronal holes dominated by open magnetic field lines and a wide streamer belt centered on the solar equator. The fine-scale magnetic morphology inferred from the white light image is showcased by the quantified topological map in the bottom left panel of Fig. \ref{fig1} (previously presented in \citealt{Boe2020b}). Further, there were not any Coronal Mass Ejections (CMEs) or flares of any kind reported on the day of the eclipse by the Space Weather Database Of Notification, Knowledge, Information (DONKI)\footnote{\url{https://ccmc.gsfc.nasa.gov/donki/}}. Hence the corona on this day is an excellent example of a minimally perturbed corona. However, just because the Sun was at solar minimum does not inherently preclude the occurrence of a CME perturbing the corona. Indeed, \cite{Boe2021b} found a large CME in the corona during the 2020 TSE, that occurred less than 18 months after the 2019 TSE presented here, which demonstrated large scale dynamical interactions between an active region CME and a nearby streamer.

\par

\begin{figure*}[t!]
\centering
\includegraphics[width = 7in]{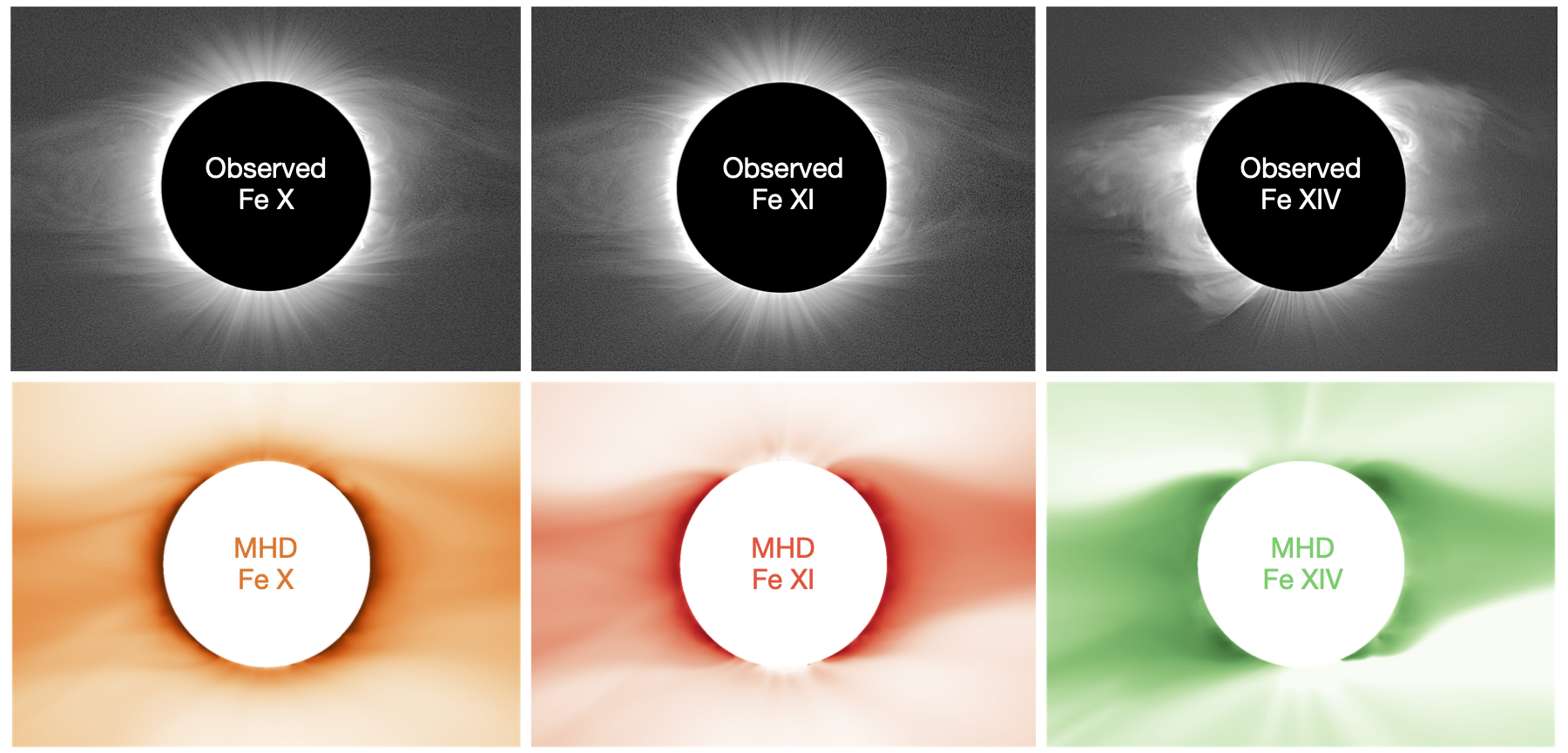}
\caption{Top: Images of \ion[Fe x], \ion[Fe xi], and \ion[Fe xiv] line emission (see \citealt{Habbal2021}), with solar north upward. All images have been processed to remove the radial gradient in brightness and enhance the fine-scale structures. Bottom: Prediction of emission from the same lines as above from the PSI MHD model. These images have been radially flattened and are shown in logarithmic scale to enhance fine-scale details in the emission (analogous to the top panels).}
\label{fig2} 
\end{figure*}

In addition to the white-light data, we present observations of the \ion[Fe x] (637.4 nm), \ion[Fe xi] (789.2 nm) and \ion[Fe xiv] (530.3 nm) emission lines. In the top panels of Fig. \ref{fig2}, we show processed versions of the \ion[Fe x], \ion[Fe xi], and \ion[Fe xiv] line emission, where the radial brightness gradient has been removed to highlight small scale spatial variations (see \citealt{Druckmuller2006}).

\par
\begin{figure*}[t!]
\centering

\includegraphics[width = 2.5in]{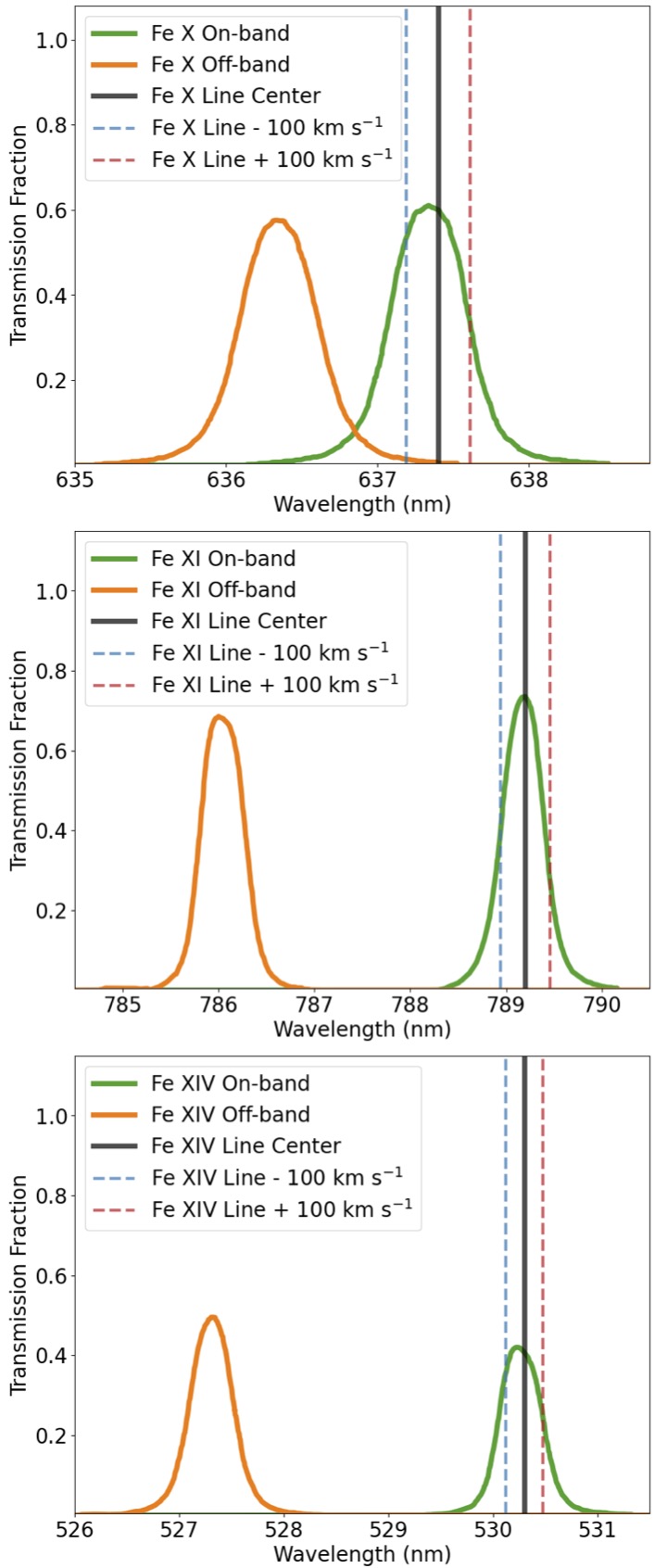}
\caption{The set of narrowband filter profiles for the line (On-band) and continuum (Off-band) of \ion[Fe x], \ion[Fe xi] and \ion[Fe xiv], as measured by the manufacturer. The bandpasses are indicated by green curves for the On-band and by orange curves for the Off-band, which are shifted to slightly lower wavelengths than the line emission. Each line center is denoted by a vertical black line along with dashed blue- and red-shifted locations at 100 km s$^{-1}$ (which is faster than expected for Doppler velocities in the corona).} 
\label{Fig3}
\end{figure*}

The line emission data were acquired with narrowband imaging systems that had bandpasses of $\approx$ 0.5 nm, which we refer to as ``On-band". For each emission line observation, we make an additional continuum observation with an identical telescope system, but with its bandpass shifted 1--3 nm toward the blue of the emission line, which were refer to as ``Off-band". The bandpasses for all emission lines (i.e. On-Band) and corresponding continuum observations (i.e. Off-Band) are shown in Figure \ref{Fig3}. The bandpass curves are from the measured transmission of the filters by the manufacturer (Andover corporation). Similar imaging systems have also been used at many previous eclipses (see \citealt{Habbal2010b, Habbal2013, Habbal2014, Habbal2021, Boe2018, Boe2020a}). Details on these telescopic systems and observational methodology for this eclipse specifically were discussed at length in \cite{Boe2021a}, where the set of Off-band observations across the visible spectrum were used to isolate the K- and F-corona using a color based inversion technique.

\par

\begin{figure*}[t!]
\includegraphics[width =2.53in]{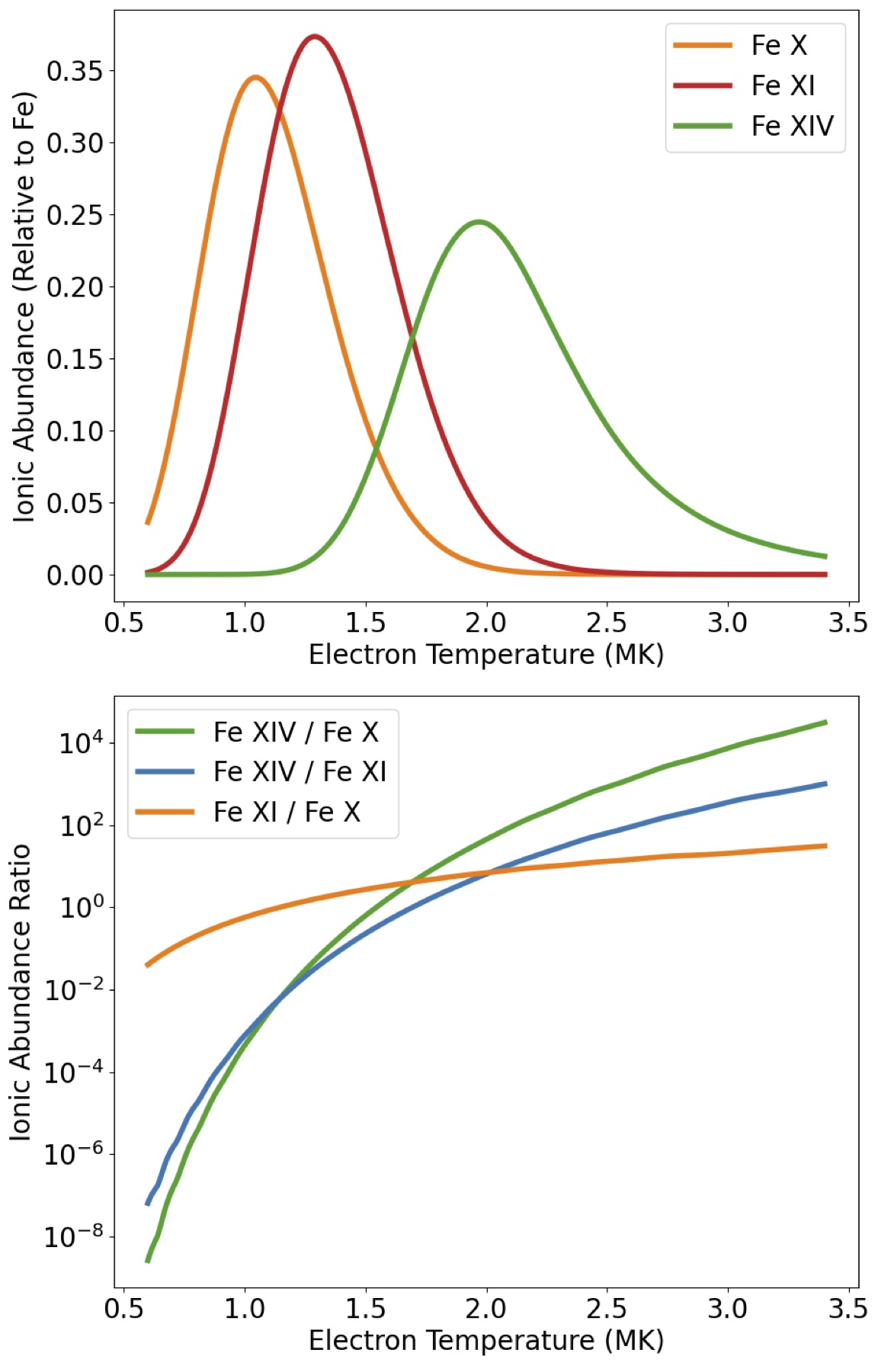}
\caption{Top: Ionic abundances versus $T_e$ with values interpolated from CHIANTI for \ion[Fe x], \ion[Fe xi] and \ion[Fe xiv]. Bottom: Ionic abundance ratios from the ionic abundance curves above versus $T_e$.}
\label{Fig4}
\end{figure*}

These emission line observations allow us to probe coronal plasmas at different $T_e$'s spanning the expected range of values from 0.8 to 2.5 MK, per the ionic abundance curves shown in the top panel of Figure \ref{Fig4}, which are taken from version 10 of the CHIANTI database \citep{Dere1997, DelZanna2021}. These ionic equilibrium data have been interpolated from their recorded spacing of $\Delta$ log(K) = 0.05 into the smooth curves shown in the figure.

\newpage
\subsection{Data Calibration}
\label{Calib}

Since the continuum data were already calibrated into solar disk brightness units ($B_\odot$, see \citealt{Boe2021a}), via the Mauna Loa Solar Observatory's (MLSO) K-coronagraph (K-Cor)\footnote{K-Cor DOI: 10.5065/D69G5JV8; \url{https://mlso.hao.ucar.edu/mlso_data_calendar.php}}, we used the Off-band to calibrate the On-band data. First, we measured the sky brightness at the center of the Moon during totality in each composite eclipse image (both On- and Off-band) and subtracted it to set the correct zero point. Next, we took the pixel flux in a region in the corner of each On- and Off-band image $>3.5$ \Rs, and set them equal to each other in each image pair. In doing so, we assumed that the line emission is negligible at that distance given the dominance of the F-corona emission over both the K corona and line emission (see \citealt{Boe2021a}). Any remaining K-corona at that distance would further wash out the presence of negligible line emission. Once the On- and Off-band pairs have been self-calibrated, we added back the expected Earthshine brightness on the moon of $2.5 \times 10^{-10}  \pm 1.5 \times 10^{-10}  B_\odot$ \citep{Agrawal2016}. Once the data were transformed into solar disk brightness units, the continuum (Off-band) images were then subtracted from the On-band data to isolate the brightness of each emission line. 

\subsection{Line width correction}
\label{LineWidth}
Finally, we performed a correction to the line emission data based on the expected line widths and known bandpasses. The bandpasses were designed to be broader than the line widths, but the bandpasses are not a perfect boxcar function (see Fig \ref{Fig3}), so we had to account for the effect that the shape of the bandpass had on the final integrated line brightness.
\par
To determine the expected line widths, we used the expected effective ion temperatures ($T_{\text{eff}}$). $T_{\text{eff}}$ is a measure directly determined by the line width according to the following equation:
\begin{eqnarray}
\label{Teff}
T^i_{\text{eff}} &&= \frac{m_i}{2k}\Delta \lambda^2 \Big( \frac{c}{\lambda} \Big)^2,\
\end{eqnarray}
where $m_i$ is the mass of the ion, $k$ is Boltzmann's constant, $c$ is the speed of light, $\lambda$ is the line wavelength, and $\Delta \lambda$ is the Doppler width, defined by the standard deviation of the Gaussian line (i.e. $\sigma$). Note that the full width at half maximum (FWHM) is $ 2.355 \ \sigma$ for a Gaussian function. Equation \ref{Teff} and an estimate of $T_{\text{eff}}$ then allows one to determine the expected width for an arbitrary line. $T_{\text{eff}}$ is related to the ion temperature, with additional broadening due to turbulent and non-thermal motions. Since we only need an estimate of the line width, and not the actual ion temperature, $T_{\text{eff}}$ is sufficient for our purposes. 
\par
\begin{figure*}[t!]
\centering
\includegraphics[width = 2.5in]{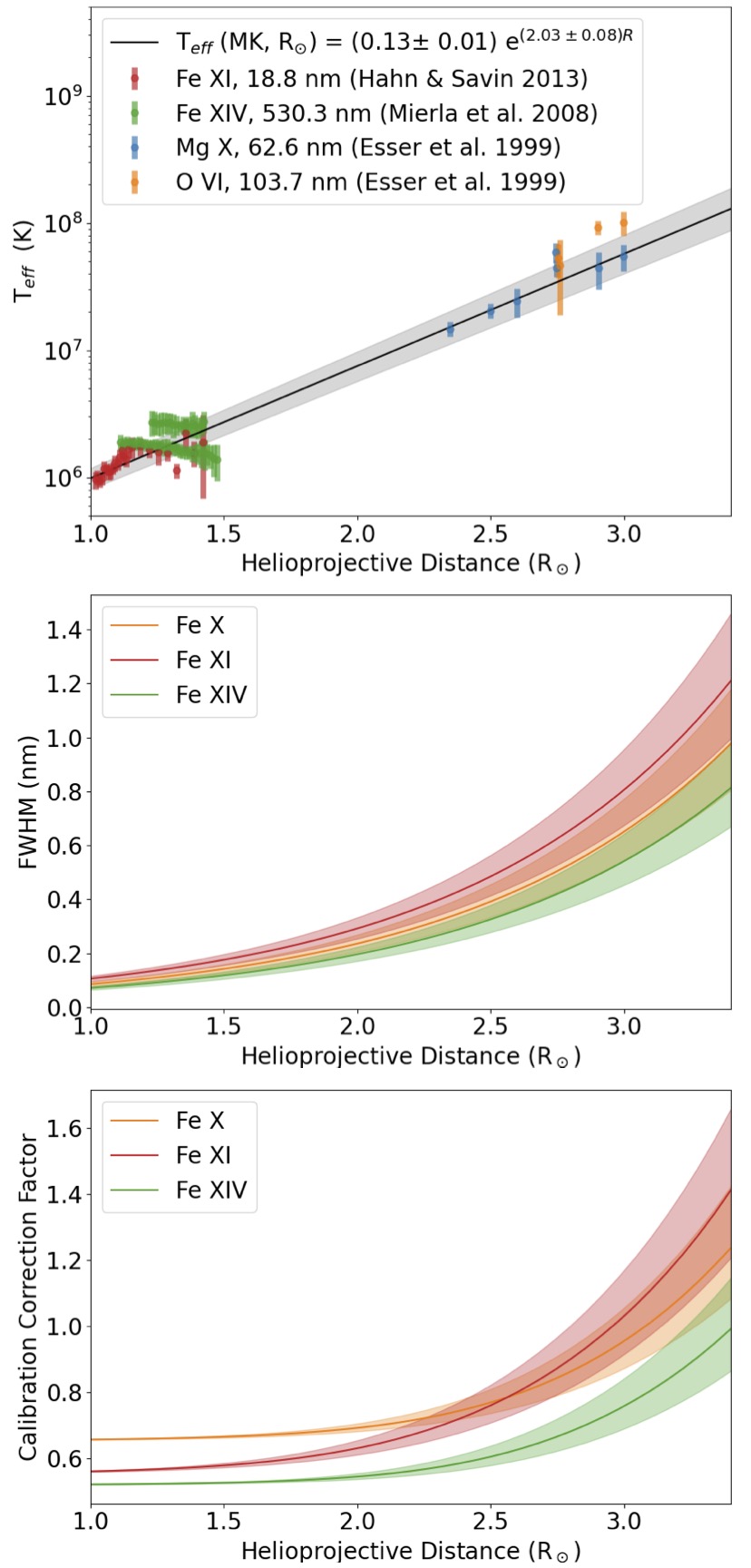}
\caption{Top: Effective ion temperature data from various spectral line observations \citep{Esser1999,Mierla2008,Hahn2013}. The best fit of the data is indicated by the black line, with the grey band indicating the uncertainty of the fit. Middle: Calculated FWHM for the \ion[Fe x], \ion[Fe xi] and \ion[Fe xiv] lines at different helioprojected distances based on the $T_{\text{eff}}$ fit. Bottom: The calibration correction factor for each line ($C_i$, see Equation \ref{bandpassInt}), given the FWHM's in the middle panel.}
\label{Fig5}
\end{figure*}

For our estimate of $T_{\text{eff}}$ throughout the corona, we used a number of previously published estimates. We used UVCS/SOHO measurements of $T_{\text{eff}}$ of the \ion[Mg x] 62.6 nm and \ion[O vi] 103.7 nm lines, as reported by \cite{Esser1999}, which probed helioprojective distances beyond 2.2 \Rs. For lower helioprojective distances, we used line width observations from the LASCO-C1 coronagraph of \ion[Fe xiv] 530.3 nm \citep{Mierla2008}, which we converted to $T_{\text{eff}}$ using equation \ref{Teff}. We also used EUV observations from EIS/Hinode observations, reported as the effective ion velocity (i.e., $v_{\text{eff}} = c \ \Delta \lambda / \lambda$) of the \ion[Fe xi] 18.8 nm line \citep{Hahn2013}. We then fit an exponential function to this set of data using Scipy {\fontfamily{pcr}\selectfont curve$\_$fit} (as shown in the top panel of Figure \ref{Fig5}), which gave $T_{\text{eff}} = (0.13 \pm 0.01) \ e ^{(2.03 \pm 0.08) R}$, with $R$ given in units of \Rs, and $T_{\text{eff}}$ in MK. With the fit to $T_{\text{eff}}$, we computed the expected FWHM for each observed emission line, as shown in the middle panel of Figure \ref{Fig5}.

\par
To determine the correction factor needed for each line, we had to determine the relative transmission of the line and continuum signals integrated over the bandpasses. The line transmission efficiency was determined as the integrated product of a Gaussian function of the expected line width with the known bandpass for each line. The efficiency of transmitting the K+F corona signal, the integral of which was calibrated to the solar disk brightness (see Section \ref{Calib}), was found by integrating a flat continuum source over the bandpass since the K+F corona continuum is effectively flat over the narrow bandpasses. The ratio of these integrals gives the correction factor for each line, $C_i$, written as:
\begin{eqnarray}
\label{bandpassInt}
C_i = \frac{\int b_i (\lambda) \ d\lambda}{\int G_{i}(\lambda, \Delta \lambda) \ b_i (\lambda) \ d\lambda}
\end{eqnarray}
where $b_i$ is the bandpass transmission function (see Fig. \ref{Fig3}) for line $i$ with a Gaussian line profile $G_{i}$ defined by the line width of $\Delta \lambda$ that will vary with helioprojective distance. This correction factor accounts for the relative transmission of the bandpasses, and converts the $B_\odot$ unit used with the K+F corona, into an analogous unit relative to the Gaussian line profiles. That is, the conversion translates the line signal into solar disk brightness units integrated over the Gaussian line profile. 
\par
As expected, in regions where the line widths are substantially smaller than the bandpass, this correction approached the limit where the line is a delta function at the center of the bandpass. As the line grows in width, it started to cover regions of the bandpass where the transmission is lower, and so the relative transmission of the line changes. Only in the outer corona beyond about 2 \Rs \ does the increasing line width start to become noticeable in the calibration correction. The rather small changes in the calibration corrections for different FWHM also strengthens the validity of our assumptions, since even rather large changes in the assumed $T_{\text{eff}}$ leads to exceptionally small changes in the calibration correction factor. At future TSEs, we intend to deploy slightly larger bandpasses (perhaps 1 nm) to avoid this behavior altogether at larger helioprojective distances. 

\section{Eclipse Observables}
Despite the large abundance of visible and near-infrared coronal line emission observations via both TSE and coronagraphs, this work is the first to quantify the absolute brightness of \ion[Fe x], \ion[Fe xi] and \ion[Fe xiv] optical emission lines beyond $\approx$ 1.7 \Rs \ (see Section \ref{intro}). In Section \ref{Line}, we discuss the spatially resolved brightness values inferred for the various emission lines, in Section \ref{Temp} we use the line ratios between the Fe lines to infer the spatially resolved coronal $T_e$. 

\subsection{Absolute Brightness of Fe X, XI and XIV}
\label{Line}

\begin{figure*}[t!]
\centering
\includegraphics[width = 7.2in]{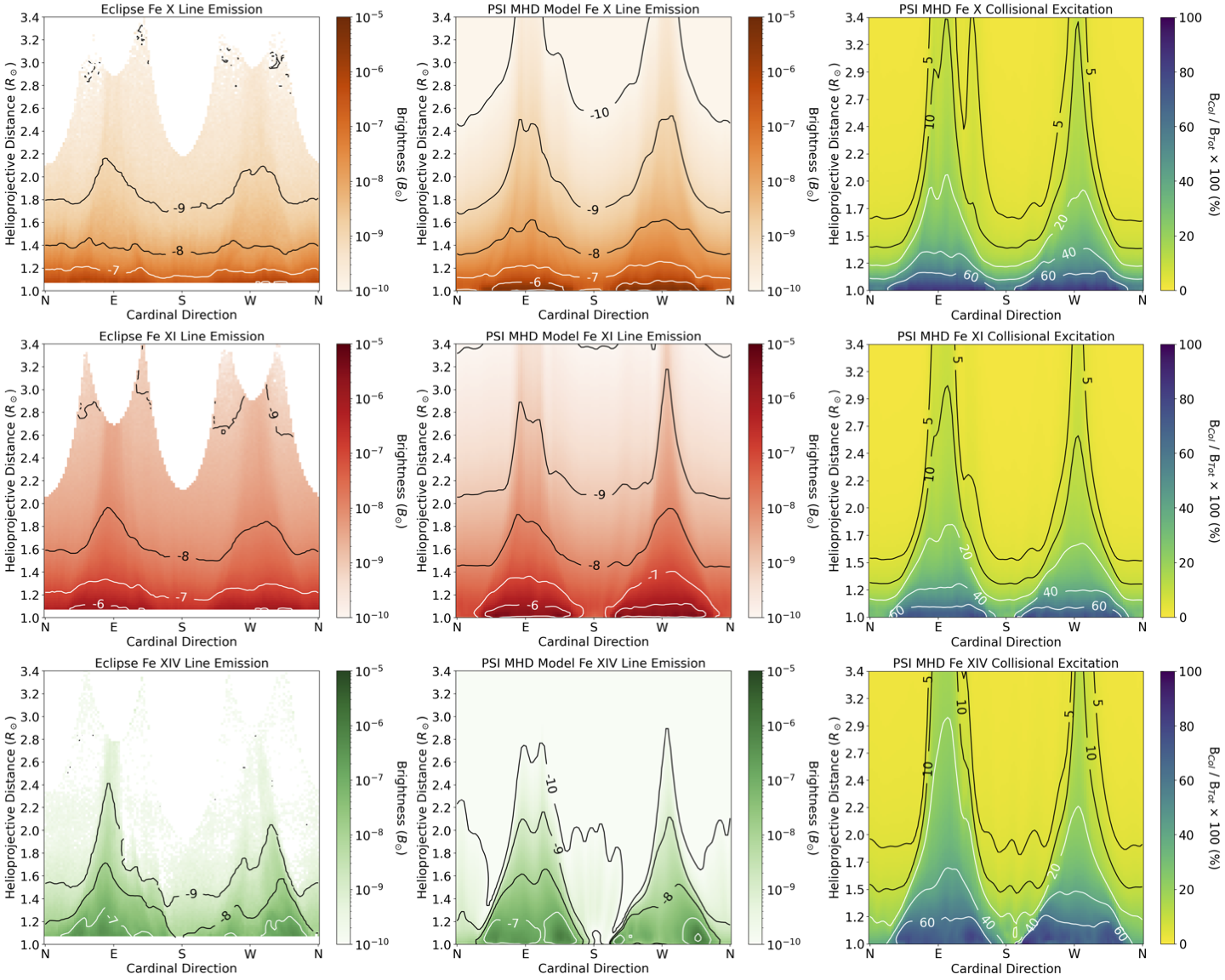}
\caption{Left panels: Calibrated absolute line emission brightness of \ion[Fe x], \ion[Fe xi] and \ion[Fe xiv]. The panels are shown in a cartesian representation of polar coordinates, with contours indicating decades of brightness. Middle panels: Same as left but for the PSI MHD prediction of the brightness for each line (see Section \ref{PSI}). Right panels: PSI MHD prediction of the percentage of each line brightness that is caused by collisional excitation.}
\label{Fig6}
\end{figure*}

The photometrically calibrated line emission data in solar disk brightness units integrated over the Gaussian line profiles (as described in Sections \ref{Calib} and \ref{LineWidth}) are shown in the left panels of Figure \ref{Fig6}. They have been sliced into a cartesian representation of polar coordinates for analysis purposes.This slicing also helps to increase the signal-to-noise (SNR) ratio in the outer corona, since higher coronal projected distances have the signal averaged from more pixels in each bin. The data is only utilized in pixels where the SNR $> 2$ of the photometric data, as determined by the propagation of photometric errors for both the On- and Off-band frames. We also overplot contours of powers of 10 of brightness in units of the solar disk brightness. 
\par

Although one could convert these brightness values into absolute units by using the solar spectrum irradiance at the wavelength of the given line, here we chose to leave the values in units of solar disk brightness. The main motivation for this choice is that these lines are radiatively excited in the corona by photospheric photons, so the absolute energy irradiated by the ions will be tied to the incoming radiation from the Sun. Since we are interested in studying the physical properties of the corona, it makes more sense to consider the lines relative to the photospheric spectrum to remove any wavelength dependent effects. Furthermore, to infer the relative ionic abundances (and thus $T_e$) one must also account for the incident photon flux in the corona, which is achieved through the solar brightness units (see Section \ref{Temp}). 

\par

\begin{figure*}[t!]
\centering
\includegraphics[width = 7in]{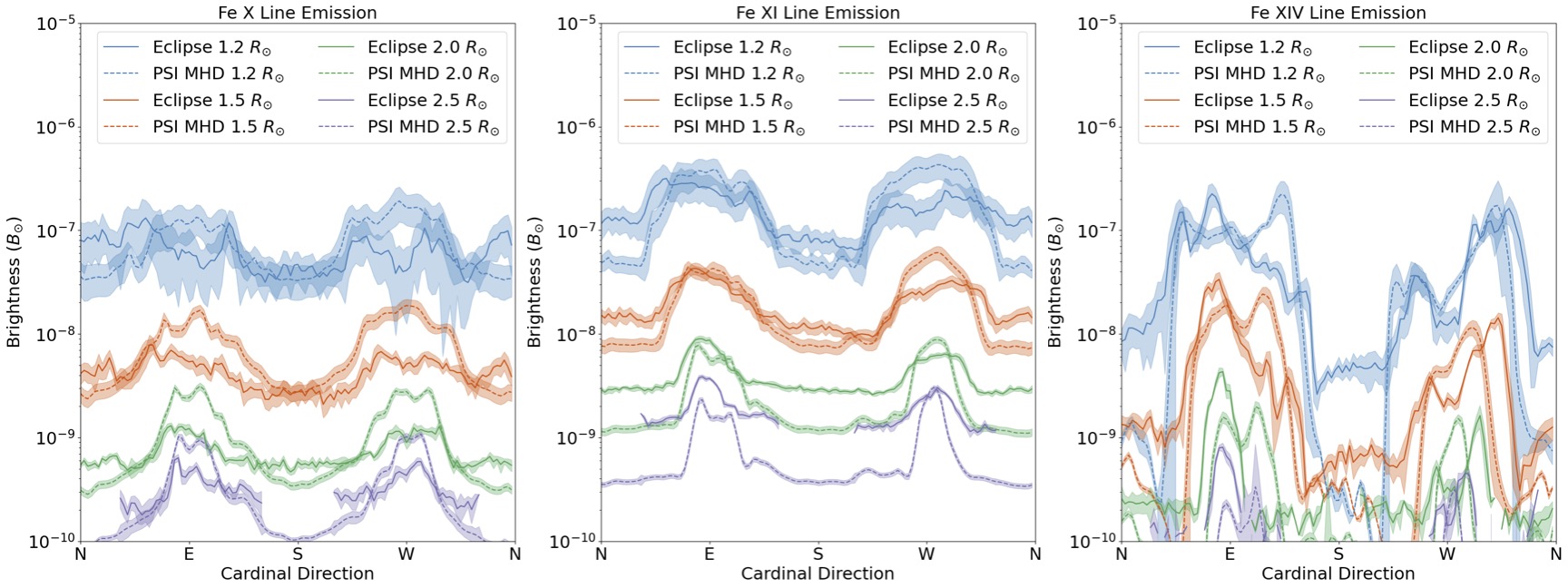}

\caption{Latitudinal traces of the brightness of \ion[Fe x] (left), \ion[Fe xi] (middle), \ion[Fe xiv] (right) from the eclipse data (solid lines) and the PSI MHD model (dashed lines). The traces are taken as the median average at fixed helioprojective distances of 1.2 (blue), 1.5 (orange), 2 (green) and 2.5 (purple) \Rs, within 0.1 \Rs \ of each distance. The filled region indicates the 1-$\sigma$ scatter of the data points used in the trace.}
\label{Fig7}
\end{figure*}

\begin{figure*}[t!]
\centering
\includegraphics[width = 7.in]{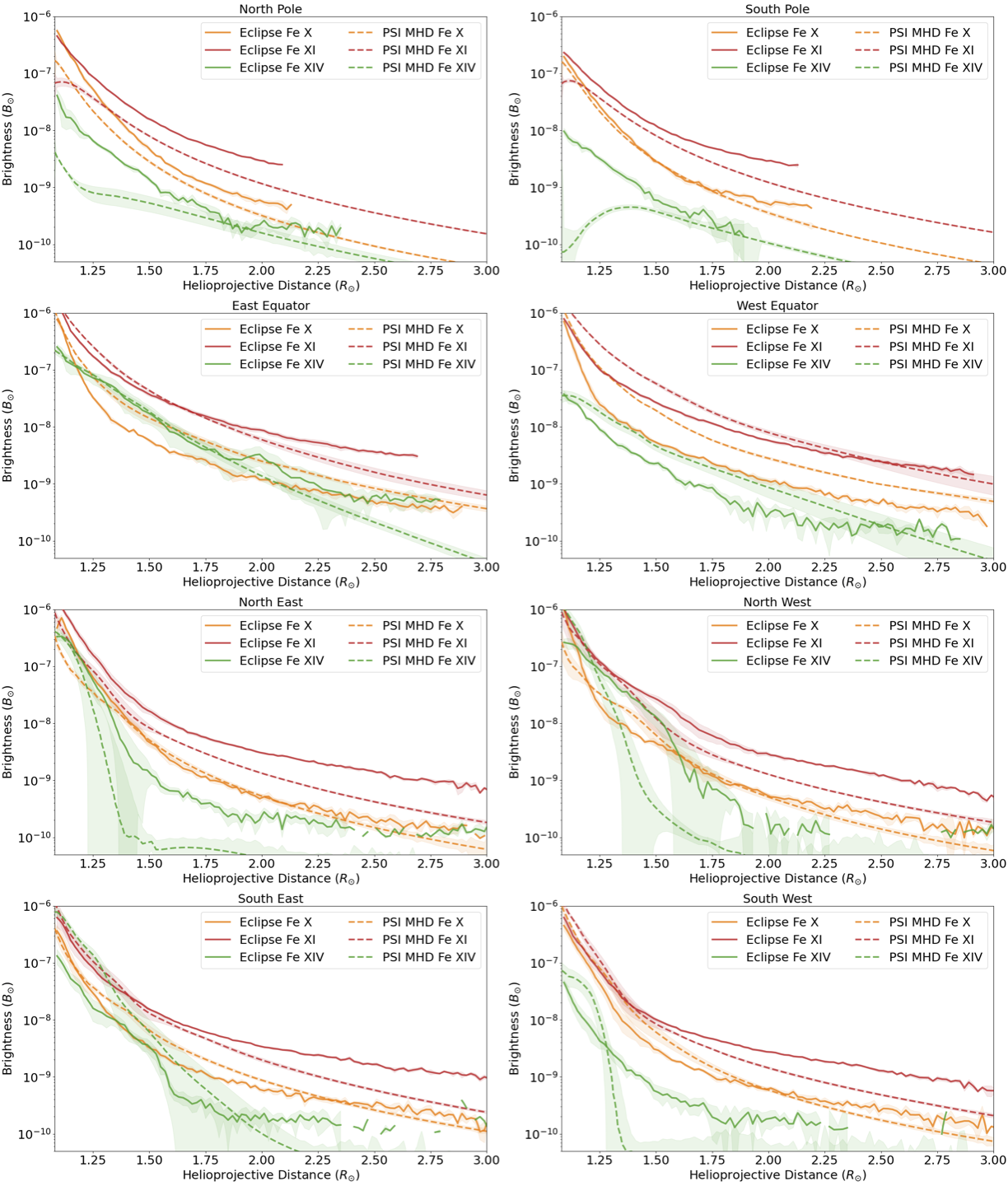}

\caption{Radial traces of the brightness of \ion[Fe x] (orange), \ion[Fe xi] (red) and \ion[Fe xiv] (green) from the eclipse data (solid lines) and the PSI MHD model (dashed lines). The traces are taken from the median average inside a 15 degree wedge centered on the cardinal direction indicated in the title of each panel. The filled bands show the 1-$\sigma$ scatter of data points within the wedge.}
\label{Fig8}
\end{figure*}

The latitudinal differences of the line emission are best illustrated in Figure \ref{Fig7}, where we show traces of the line brightnesses at the fixed helioprojective distances of 1.2, 1.5, 2, and 2.5 \Rs. The radial drop off of the line brightness is most evident in Figure \ref{Fig8}, where radial traces of the line emission are shown for a set of latitudinal regions.

\par

It is clear from both Figures \ref{Fig7} and \ref{Fig8} that \ion[Fe xi] is by far the brightest emission line of the three. In fact, it is the only line which retains high signal-to-noise ratio data until 3.4 \Rs, with a brightness of about $5 \times 10^{-10}  B_\odot$. The extent of the \ion[Fe xi] emission data was limited more by the size of the detector used for the observations than by the strength of the line emission. In the future, we plan to use a wider field of view with a larger detector to enable observations of line emission out to even higher helioprojective distances. 
\par
In streamers, the brightness of \ion[Fe xi] ranges from just over $10^{-6} B_\odot$ below 1.2 \Rs, and falls to about $5 \times 10^{-9} B_\odot$ by 3 \Rs. The spatial distribution of \ion[Fe xi] emission roughly correlates with the K-corona (electron scattering) emission inferred for this eclipse (see \citealt{Boe2021a}). The brightness of the K corona is 2 to 3 times brighter than \ion[Fe xi] throughout the corona, but the line brightness drops to only a small fraction (about $10\%$) of the total continuum K + F corona brightness beyond about 2 \Rs.

\par
The \ion[Fe x] emission behaves somewhat similarly to \ion[Fe xi], albeit with a considerably lower brightness. It reaches only about $5 \times 10^{-7} B_\odot$ below 1.2 \Rs, and fades to about $10^{-10} B_\odot$ by 3 \Rs. The \ion[Fe x] emission also has a much less pronounced variation with solar latitude, especially at distances below $\approx 1.5$ \Rs, where the brightness of the streamers and coronal holes is comparable. Above 1.5 \Rs, on the other hand, the brightness variation of \ion[Fe x] looks much more similar to \ion[Fe xi] shifted down by an order of magnitude.
\par
By contrast, the \ion[Fe xiv] emission shows the largest difference between coronal holes and streamers, with fine-scale variations throughout the streamers. It is almost as bright as \ion[Fe xi] in the core of streamers below 1.5 \Rs, but fades very quickly at higher elongations. Indeed, \ion[Fe xiv] is the only Fe line that is virtually undetectable above the noise in some regions as low as 2 \Rs, while maintaining a higher brightness in streamers out to as much as 2.8 \Rs. Since \ion[Fe xiv] emission originates from higher temperature plasmas ($T_e >$ 1.5 MK, see Fig. \ref{Fig3}), the brightness variation implies a very low proportion of high temperature plasmas at the poles of the Sun and outside the cores of streamers. 
\par

\newpage

\subsection{Electron Temperature}
\label{Temp}

The absolutely calibrated Fe line emission from all three lines enables the inference of the density weighted average $T_e$ for each LOS in the corona by comparing their relative brightness under the assumption of photoexcitation alone. These optical emission lines do have a component of collisional excitation in the low corona below $\approx$ 1.2 \Rs, as shown analytically by \cite{Habbal2007}, and for this eclipse specifically via the PSI MHD model (see Section \ref{PSI} and Fig \ref{Fig6}). To correct for the collisional emission, we multiply each emission line by the fractional amount of light that is expected to be from radiative excitation alone (from the MHD model). Since we will use the line ratio to infer $T_e$, this correction is somewhat independent of the exact density in the model and rather corrects for the relative intrinsic sensitivity that the specific emission line has to collisions.
\par
Here we apply the methodology of \cite{Boe2020a}, who used the calibrated ratio of \ion[Fe xiv]/\ion[Fe xi] from data acquired at multiple sites during the 2017 TSE, and expand it to line ratios including \ion[Fe x].

\par
A given ionic abundance ratio ($n_j/n_k$) can be related to the intensity ratio of line emission $I$, for two lines $j$ and $k$ as in the following equation:
\begin{eqnarray}
\label{ionRat}
\frac{n_{j}}{n_{k}} = \frac{I_j \ \rho(\nu_{k}) \ \epsilon_{k} \ A_{k}  \ g_{l,j} \ g_{u,k} \ {\nu_{k}}^3}{I_k \ \rho(\nu_{j}) \ \epsilon_{j} \ A_{j} \ g_{u,j} \ g_{l,k} \ {\nu_{j}}^3},
\end{eqnarray}
where the $A$ values are the Einstein coefficients for spontaneous emission, $g_u$ and $g_l$ are the statistical weights for the higher and lower energy level of the given transition, $\nu$ is the frequency of the light emitted by the transition, the $\epsilon$ values indicate the photometric efficiency of the telescope systems (and Earth's atmosphere), and the $\rho(\nu)$ terms are the volumetric photon energy density in the corona where the ions are being excited. The $A$, $g$ and $\nu$ constants used for this work are shown in Table \ref{table2}.
\par

\begin{table}[t]
\begin{center}
\begin{tabular}{ c  c  c  c  c c }
\hline
Line &  $\lambda_{ion}$ (nm) & $\nu_{ion}$ ($10^{14}$ Hz)  & $A_{ion}$ ($s^{-1}$) & $g_l$ & $g_u$  \\
\hline
 \ion[Fe x]  & 637.5 & 4.70 & 69.4 & 4 & 2  \\
\hline
 \ion[Fe xi]  & 789.2 & 3.80 & 43.7 & 5 & 3  \\
\hline
 \ion[Fe xiv]  &530.3 & 5.65 & 60.2 & 2 & 4  \\
\caption{Constants used for equation \ref{eqnFinal}. Data from NIST (\citealt{NIST_ASD} and ref. therein)}
\label{table2}
\end{tabular} 
\end{center}
\vspace{-9mm}
\end{table}

Equation \ref{ionRat} can be simplified since the product $\rho(\nu) \ \epsilon$ for each line is accounted for in the absolute photometric calibration of the line brightness into solar disk brightness units (see Section \ref{Line}). The $\rho(\nu)$ terms are technically composed of the solar spectral energy distribution (accounted for via the solar disk brightness unit) and the volumetric scattering geometry of the corona and extended solar disk. However, these volumetric considerations will be identical for any two lines integrated over the same LOS, hence they disappear in a line ratio. Equation \ref{ionRat} then reduces to:
\begin{eqnarray}
\label{eqnFinal}
\frac{n_{j}}{n_{k}} = \frac{\beta_j \ A_{k}  \ g_{l,j} \ g_{u,k} \ {\nu_{k}}^3}{\beta_k \ A_{j} \ g_{u,j} \ g_{l,k} \ {\nu_{j}}^3},
\end{eqnarray}
where the $\beta$ terms refer to the brightness of the emission lines in units of the solar disk brightness, rather than the absolute intensity of the lines (so $\beta_i = I_i \ \rho(\nu_i)^{-1} \ \epsilon_i^{-1} $). 
\par
\begin{figure*}[t!]
\centering
\includegraphics[width = 7in]{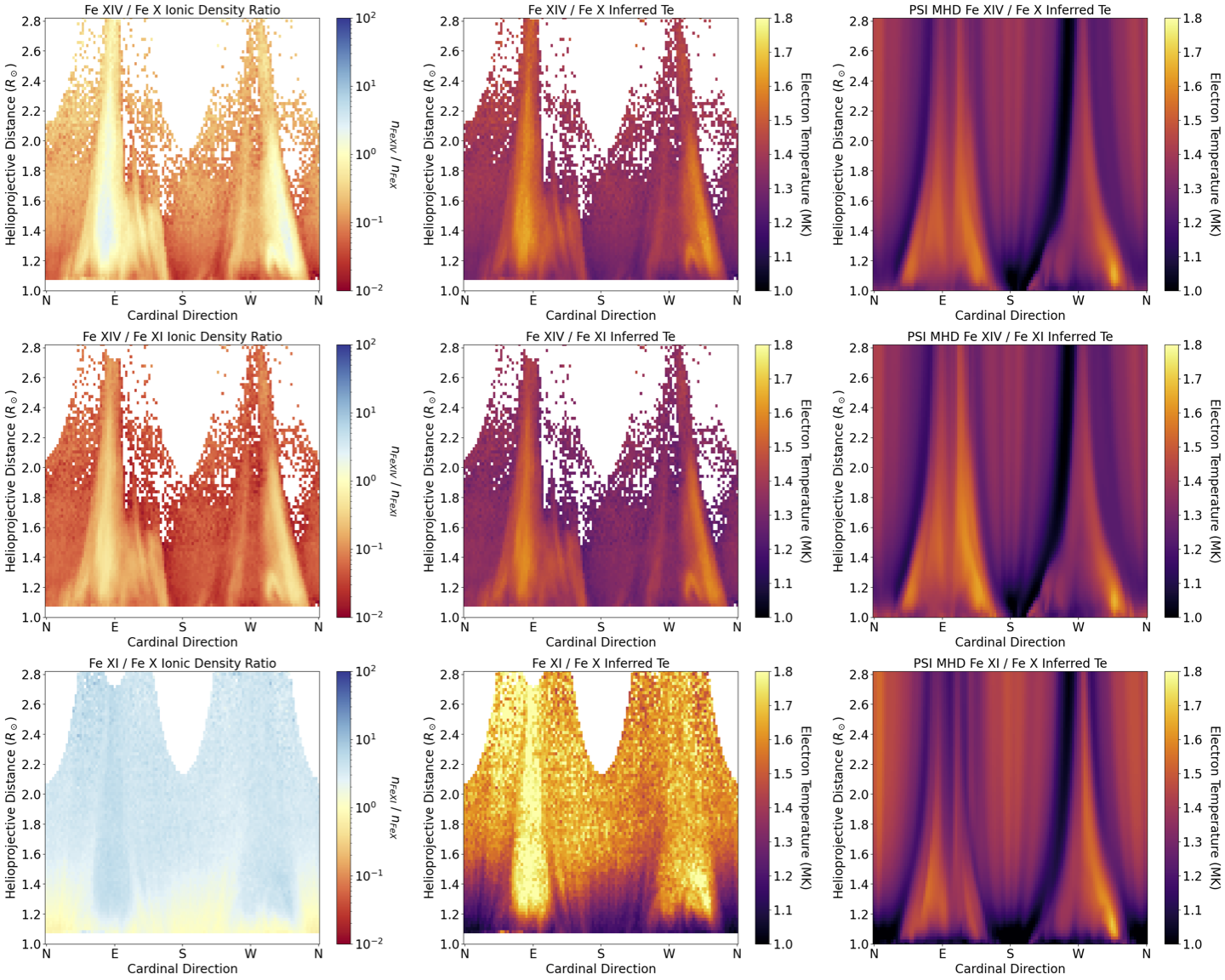}
\caption{Left panels: Ionic density ratios inferred from the line ratios of \ion[Fe xiv]/\ion[Fe x] (Top), \ion[Fe xiv]/\ion[Fe xi] (Middle), and \ion[Fe xi]/\ion[Fe x] (Bottom) using Equation \ref{eqnFinal} and the observed line brightnesses. Middle panels: Inferred $T_e$ from the ionic density ratios to the left, using the ionization ratio curves shown in Figure \ref{Fig4}. Right panels: Inferred $T_e$ using the predicted line emission from the PSI MHD model (see Section \ref{PSI}).}
\label{Fig9}
\end{figure*}

Next, we convert the ionic density ratio from Equation \ref{eqnFinal} to an inference of $T_e$ based on the ionic abundances as a function of $T_e$ from CHIANTI (see Section \ref{Eclipse} and Fig. \ref{Fig3}). With the \ion[Fe x], \ion[Fe xi] and \ion[Fe xiv] observations, we calculate three different line ratio temperatures for each possible combination. The resulting ionic density ratios and $T_e$ maps are shown in Figure \ref{Fig9}. We only display pixels where the SNR $>2$ for both lines used in each line ratio $T_e$ inference. These panels are shown out to a heliocentric distance of 2.8 \Rs, rather than 3.4 \Rs \ as in Figure \ref{Fig6}. The lower distance was used for this figure because the signal of the \ion[Fe xiv] line specifically was exceptionally weak beyond this distance, and so the inferred line ratios would not be robust for distances beyond 2.8 \Rs.

\par

We find that Fe $10^+$ (\ion[Fe xi]) is the most abundant ion throughout the corona, even in higher $T_e$ regions. Fe $9^+$ (\ion[Fe x]) is comparable to Fe $10^+$ in the low corona below 1.3 \Rs, but is 3 to 5 times less abundant beyond that distance. Fe $10^+$ is always more abundant than Fe $13^+$ (\ion[Fe xiv]) everywhere in the corona. In fact, Fe $13^+$ is never more than $\approx 50\%$ as abundant as Fe $10^+$, even in the hottest regions at the core of the Eastern streamer. Nevertheless, Fe $13^+$ is up to 2 times as abundant as Fe $9^+$ in the high $T_e$ regions. The Fe $13^+$ abundance drops dramatically in open field regions however, often to less than $10\%$ the abundance of Fe $9^+$. The result that Fe $10^+$ is the most abundant ion in the corona supports \cite{Habbal2010a, Habbal2021}, who found it to be the most abundant Fe ion in the solar wind and is the brightest Fe emission line in the corona, regardless of solar cycle.

\par

\begin{figure*}[t!]
\centering
\includegraphics[width = 7in]{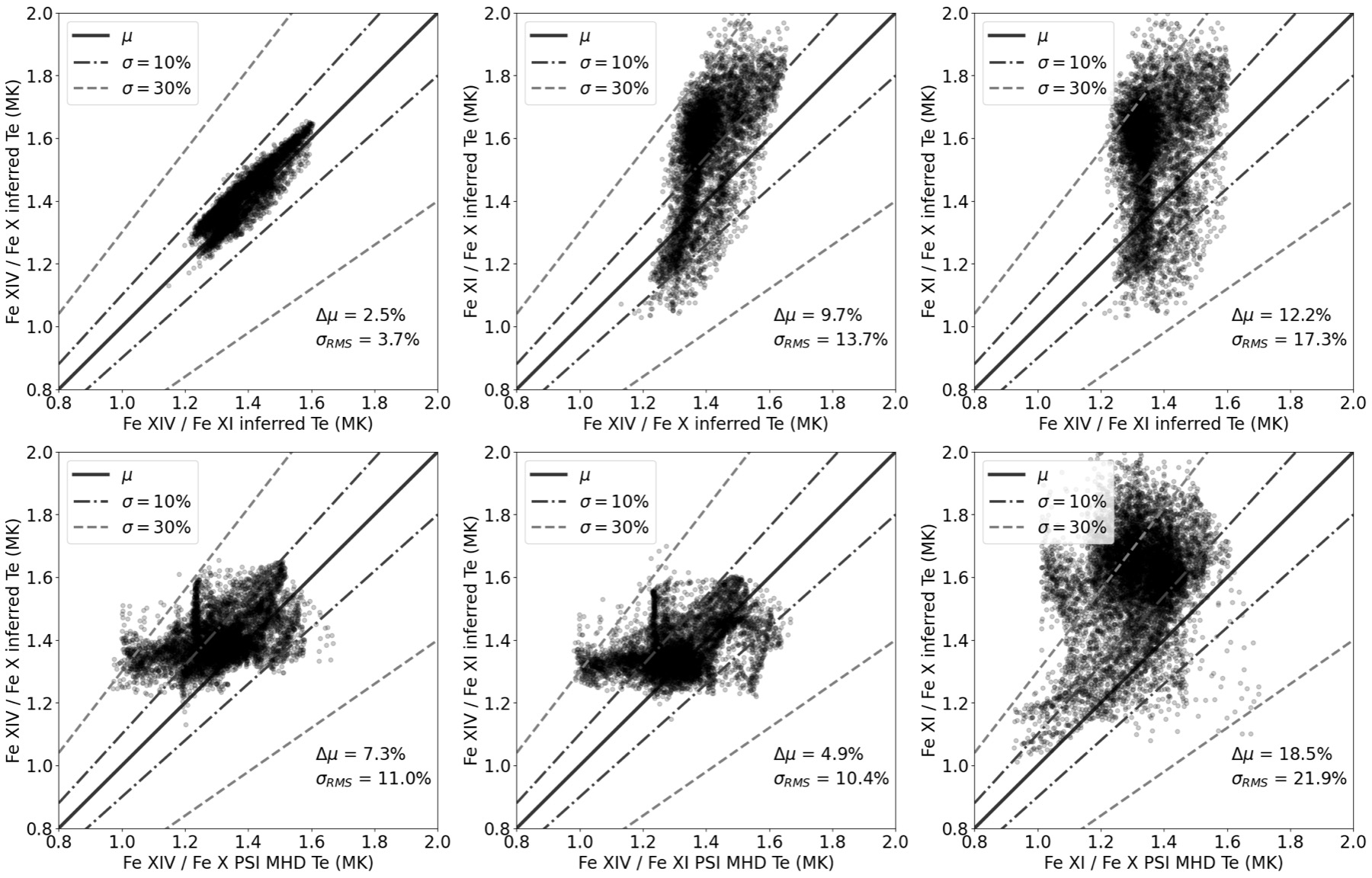}
\caption{Comparative scatter plots of the electron temperature from the \ion[Fe xiv]/\ion[Fe x] ,  \ion[Fe xiv]/\ion[Fe xi], and  \ion[Fe xi]/\ion[Fe x] line ratios (see Section \ref{Temp}), along with the PSI MHD prediction driven inferences from the same line ratios (see Section \ref{PSI}). The solid line shows the 1-1 correspondence, the dot-dash (dashed) lines show a $\pm 10 \%$ ($\pm 30 \%$) variance from the 1-1 correspondence. The mean offset is written as $\Delta \mu$ with the sign indicating whether the average is above or below the 1-1 line, and the $\sigma_{RMS}$ indicates the root-mean-square variance of the data points around the mean offset.}
\label{Fig10}
\end{figure*}

The $T_e$ values inferred from the \ion[Fe xiv]/\ion[Fe xi] and \ion[Fe xiv]/\ion[Fe x] line ratios are consistent with each other, as demonstrated by the direct comparison between these $T_e$ inferences for each LOS in the top left panel of Figure \ref{Fig10}. The \ion[Fe xiv]/\ion[Fe x] inferred $T_e$ is only 2.5$\%$ higher than the \ion[Fe xiv]/\ion[Fe xi] $T_e$ value on average. The root-mean-square (RMS) variance of the average is 3.7 $\%$, so the two methods are statistically equivalent. Both inferences range from about 1.25 -- 1.4 MK in the coronal holes and about 1.5 -- 1.65 MK in the equatorial streamers. The streamers show a large $T_e$ spatial variability where the streamer cores are hotter (about 1.6 MK) than the boundaries (1.5 MK). There are also pockets of lower temperature plasmas (1.3 MK) at the base of the streamers below 1.2 \Rs. The thickness of the streamers (as visualized by the $T_e$ spatial distribution) also decreases at larger helioprojective distances, while the coronal holes dominate more of the corona with a roughly uniform temperature. 

\par

The \ion[Fe xi]/\ion[Fe x] inferred $T_e$ is quite different from the two line ratios involving \ion[Fe xiv], where it infers a much larger range of temperatures. This line ratio indicates that the coronal hole $T_e$ is closer to 1.1 MK with distinct fine-scale plumes of cool plasma emerging into the corona. The core of the streamers then is over 1.8 MK in this inference, while the outer corona is quite a bit higher in temperature than the other line ratio inferences. While the \ion[Fe xi]/\ion[Fe x] ratio is an interesting map for showcasing small temperature variations in the coronal hole plumes, it is not a very reliable measure of the temperature. One reason for this large spread of $T_e$ is that the slope of the \ion[Fe xi]/\ion[Fe x] ionic abundance ratio as a function of $T_e$ is considerably shallower than the other line ratios (see Fig. \ref{Fig4}). Therefore, a small change in the emission ratio leads to a large change in the inferred temperature. The temperature response range of the \ion[Fe x] and \ion[Fe xi] curves alone also does not probe the range above about 1.5 MK effectively; consequently this line ratio is unreliable anywhere outside the coolest regions in the corona. The large difference between the \ion[Fe xi]/\ion[Fe x] and the other line ratio $T_e$ inferences perhaps implies that the corona is not isothermal along a single LOS, and so the LOS average is not identical for different line ratio inferences. Additionally, average $T_e$ inferences from two line ratios may be biased toward finding temperatures in-between the peak ionization of the temperature response functions (i.e. Fig. \ref{Fig4}) as demonstrated in the EUV by \cite{Weber2005}. We intend to explore a more complex $T_e$ inference using all three (or more) of these visible emission lines simultaneously in a future study, as it is beyond the scope of this work.

\newpage
\section{The PSI MHD Model}
\label{PSI}
The spatially resolved coronal line emission (Fig. \ref{Fig6}) and inferred electron temperature (Fig. \ref{Fig9}) are the first such data to span the distance of 1.08 to 2.8 (up to 3.4 \Rs \ for \ion[Fe xi]), and offer a unique opportunity to test and constrain advanced models of the corona. Specifically, we compare our observational eclipse results to forward modeled line emission using the state-of-the-art Magnetohydrodynamic (MHD) simulation of the PSI/MAS model \citep{Mikic2018} for the 2019 July 2 Total Solar Eclipse corona\footnote{\url{http://www.predsci.com/corona/jul2019eclipse/home.php}}. There exists a rich heritage of forward modeling white-light and extreme ultraviolet (EUV) observables from the PSI MHD model (e.g. \citealt{Mikic1999, Mok2005, Lionello2009}), where the local emissivity for a given observable (based on the local plasma state and/or geometry) was obtained and then integrated along a given LOS through the model. We follow a similar approach to model the \ion[Fe x], \ion[Fe xi], and \ion[Fe xiv] emission lines (see Section \ref{forwardModel}).
\par
This MHD model was used by \cite{Boe2021a}, where the forward modeled K corona emission (top right panel of Figure \ref{fig1}) was compared to the eclipse data (see \cite{Boe2021a} for the details of the model). The model was found to accurately predict the brightness of the K corona, and consequently the electron density. Further, the magnetic morphology of the model (bottom right panel of Figure \ref{fig1}) closely matches the fine-scale striations seen in the eclipse image. Given that the MHD model has generated a reasonable electron density and magnetic field morphology prediction throughout the corona, we can reliably extend the comparison to include the line emission observables, which offers a unique opportunity to test other plasma parameters in the model.

\subsection{Forward Modeled Line Emission}
\label{forwardModel}
\par
Assuming local thermodynamic equilibrium, the total line emissivity of a parcel of plasma depends on the local electron density, temperature, and radiation field near line center, originating from both radiative and collisional excitation. To tackle this problem, we use the CHIANTI 10 database and software package (as described in Section \ref{Eclipse}) and its framework for adding photo-excitation to the level population calculations \citep{Young2003}. The contribution function for each Fe line individually is computed on a 3D grid of density, temperature, and solar radius. The radial distance is then used to approximate the local radiation field at these wavelengths by using the observed solar spectral irradiance from the International Space Station \citep{Meftah2018}. This lookup table for the contribution function is then used to calculate the local emissivity at each point along the LOS in the forward modeling computation. 
\par
For the abundance factor we adopt the ``hybrid" coronal abundances \citep{Schmelz2012}, which match the radiative loss function used in the thermodynamic MHD calculation. We can also turn off the photoexcitation flags in the CHIANTI calculation to compute contribution functions based on collisions alone and use these in a separate forward modeling experiment. This enables us to compare the relative importance of collisional excitation for each LOS in the forward modeled line emission. Finally, to compare the model calculations with the eclipse observations, we convert the LOS integrated emission into solar disk brightness units by using the solar spectral irradiance and accounting for the angular size of the solar disk, which had a radius of about 943 arc-seconds at the time of the eclipse (Earth was at aphelion). 
\par
Radially flattened versions of the model predicted \ion[Fe x], \ion[Fe xi], and \ion[Fe xiv] line emission are shown in the bottom panels of Figure \ref{fig2}. The photometric brightness predictions of all three lines are shown in the middle panels beside the observed emission in Figure \ref{Fig6}, with the same cartesian representation of polar coordinates as the eclipse data in the left panels. In the right panels of the figure, we show the percentage of the emission that originates from collisional excitation compared to the total brightness of each line. Collisional excitation is important in the low corona inside streamers, but the photoexcitation becomes the dominant excitation process everywhere beyond 1.2 \Rs. We use the modeled fraction of emission that is radiative to correct the emission observations specifically for the line ratio inferred $T_e$ values (as done in Section \ref{Temp}). However, it is possible that our $T_e$ inferences are not as robust at the base of streamers. This region is commonly explored through EUV observations, since EUV emission is predominantly collisionally excited, so it is not crucial for these eclipse observations to robustly probe the collisionally dominated regions of the corona. Regardless, our $T_e$ results are generally consistent with the values inferred via EUV observations during periods close to solar minimum (e.g., \citealt{Morgan2017}).

\subsection{Testing the Model}
\subsubsection{Line emission Comparisons}
While the PSI MHD model generally makes reasonably accurate predictions of the overall brightness and structure of coronal line emission as inferred from the eclipse data, there are some notable differences. These differences are best seen in Figures \ref{Fig7} and \ref{Fig8}, in the comparison of the latitudinal and radial traces of the line emission from both the eclipse data and the model. 
\par
The prediction of \ion[Fe x] is the best overall. For this line, the model and data are an excellent match below about 1.5 \Rs, and disagree only slightly for larger elongations. At the higher distances the model brightness is too high in the streamers by about a factor of 2. In the eclipse data, the \ion[Fe x] line emission is more uniform throughout the corona regardless of the coronal structures than is predicted by the model. 

\par
The prediction of \ion [Fe xi] is also quite good, especially in the streamers. However, the model significantly underestimates the \ion [Fe xi] emission in the coronal holes and the difference grows for larger helioprojective distances. This difference in coronal holes is an indication that the model is underestimating the density of \ion[Fe xi] in the outflowing solar wind. 
\par
The \ion[Fe xiv] line is more complex to compare, in large part because the line is sensitive to the higher temperature regions in the corona that are controlled by complex dynamics in the closed field regions of streamers and active regions. Indeed, the precise brightness differences on small spatial scales between the model and observations for \ion[Fe xiv] specifically are not always meaningful, since small variations are easily caused by slight errors in the angle of streamers in the model which can be tilted by quite small changes in the polar magnetic fields used in the model \citep{Riley2019}. For \ion[Fe xiv] it is then not useful to compare each individual LOS between the model and the eclipse data necessarily, but rather to compare the general trends and range of values seen from different coronal structures at various elongations. Still, the overall PSI MHD prediction of \ion[Fe xiv] is reasonably close for streamer regions, but has some large differences in the coronal holes, where the model substantially underestimates the brightness -- in a similar manner to the \ion[Fe xi] prediction.

\subsubsection{Inferred $T_e$ Comparisons}
\label{modelTe}
Next, we use the model line emission of the Fe lines to infer $T_e$, replicating the approach used with the eclipse data (see Section \ref{Temp}). We use the inferred $T_e$ from the model line emission rather than the actual $T_e$ in the model since the resulting inference will be systematically the same as the eclipse inference. The comparison between these inferences is a more direct test of the modeled temperature and density distributions, because it uses the emission line observables directly (and thus the formation mechanism, density and temperature along the LOS) as opposed to comparing directly to the plasma temperature in the model. 
\par

The PSI MHD model predictions of $T_e$ from the \ion[Fe xiv]/\ion[Fe xi], \ion[Fe xiv]/\ion[Fe x], and \ion[Fe xi]/\ion[Fe x] model line ratios are shown in the right panels of Figure \ref{Fig9}. The eclipse and model inferences for each LOS are compared in Figure \ref{Fig10}. Although the model predictions show similar qualitative structures as the eclipse inferred $T_e$ maps, the mode inferences are slightly lower on average. However, the spread of $T_e$ between the eclipse and model inferences indicate that the global average $T_e$ of the model inferences are rather similar to the eclipse inference. Specifically the \ion[Fe xiv]/\ion[Fe x] inference is 7.3 $ \pm 11.0 \%$ lower, and the \ion[Fe xiv]/\ion[Fe xi] inference is 4.9 $ \pm 10.4 \%$ lower than the eclipse inference. 
\par
As discussed in Section \ref{Temp} the \ion[Fe xi]/\ion[Fe x] line ratio is a somewhat unreliable measurement of the higher temperatures in the corona. It is then not surprising that the model and the eclipse inferences diverge for this line ratio $T_e$ for temperatures higher than about 1.4 MK. Though, the model and the data match reasonably well for the lower temperatures.
\par

These comparisons illustrate some of the ways in which these unique observables can be used to test and validate coronal models in the low- to mid-corona. The relative agreement of the model and observations in the closed field regions suggests that it is doing a reasonably good job at reproducing both the electron density and temperature distributions there, which are set by the interplay of the 3D closed magnetic field geometries and the coronal heating model. Nevertheless, the model somewhat overestimates the \ion[Fe x] brightness in the streamers beyond 1.2 \Rs, suggesting it may be slightly underestimating the $T_e$ in the outer regions of streamers.

\par
Further, the model underestimates \ion[Fe xi] and \ion[Fe xiv] in the polar coronal holes, which is particularly interesting from the perspective of constraining future models. Perhaps the simplest explanation would be that the current model is underestimating the density profiles within the coronal holes. However, the K corona analysis of \cite{Boe2021a} for this eclipse found that the model had a virtually identical K corona brightness to that inferred from observations, and the K corona brightness depends only on the 3D electron density distribution. Ruling out density, the next possibility would be a temperature effect, i.e. the model is under-predicting the average $T_e$ in this region or below, which would shift the Fe charge state distribution away from \ion[Fe xi] and \ion[Fe xiv].
\par

A temperature disagreement, in turn, could imply a deficiency in the coronal heating formalism in the MHD model, which in this case is set by a Wave-Turbulence-Driven (WTD) approach \citep{Lionello2014, Downs2016, Mikic2018}. On the other hand, the independent density constraint set by the K corona observables suggests that simply increasing heating to raise the coronal temperature may not be a viable solution, as this will greatly influence the density distribution along the flux tube as well. Instead, this may point to an issue with the distribution of heating along open flux-tubes in the low corona or in the complexity of the equations solved, such as solving for a single temperature (as is done here) or using a multi temperature and/or fluid approximation (e.g. \citealt{vanderHolst2022}).
\par

This analysis is also complicated by the fact that charge-state freeze-in is likely to have occurred relatively low in the corona (roughly 1.2 -- 2 \Rs), as supported by our analysis for the 2015 total solar eclipse \citep{Boe2018} and by empirical modeling \citep{Gilly2020}. Considering that the frozen-in charge state distributions are sensitive to the temperature, density, and velocity distributions where freeze-in occurs (e.g. \citealt{Lionello2019}), the combination of broadband K corona (density) and photo-excited emission lines (charge-states) can provide tight constraints on coronal and solar wind models. We will explore improvements to the MHD model and the role of non-equilibrium ionization on the forward modeled observables in a future study.

\section{Conclusions}
\label{conc}

The 2019 Total Solar Eclipse was an excellent example of a solar minimum corona which presented a unique opportunity to study the properties of a relatively quiescent corona. In this work, we expanded on the continuum dataset of \cite{Boe2021a} by analyzing the line emission observations of \ion[Fe x], \ion[Fe xi] and \ion[Fe xiv] which corresponded to the continuum bandpasses in that paper (see Section \ref{Eclipse}). We have presented absolutely calibrated (in units of solar disk brightness) spatially resolved line emission for these lines from 1.08 up to as much as 3.4 \Rs \ for the first time (see Section \ref{Line}). This unique combination of broadband and narrowband line observables provides strong constraints for the model and the distribution of heating as a function of height. Such constraints can be used to improve our understanding of the physics of coronal holes and the nascent solar wind.

\par
Equipped with the calibrated line emission, we inferred the electron temperature ($T_e$) using two-line ratios between each pair of Fe lines (see Section \ref{Temp}). We find that:
\begin{enumerate}
\item The Fe$^{10+}$ ion is the most abundant of the three throughout the corona, supporting the work of \cite{Habbal2010a,Habbal2021}. 
\item The \ion[Fe xiv]/\ion[Fe x] and \ion[Fe xiv]/\ion[Fe xi] line ratios produce effectively the same $T_e$ with the streamers having a temperature of about 1.65 MK at the core, closer to 1.5 MK at their border, whereas the coronal holes are at about 1.25 to 1.4 MK. 
\item The streamers have pockets of lower $T_e$ plasmas at their base from 1.08 to 1.3 \Rs, but have significantly higher $T_e$ in their cores at higher helioprojective distances.
\item The width of the streamers, as inferred from the $T_e$ distribution, reduces substantially with distance from the Sun while the rest of the corona is almost isothermal everywhere beyond 1.2 \Rs \ or so.
\end{enumerate}

\par
We then compared our line emission and $T_e$ results to the MHD model from Predictive Science Inc. The forward modeled line emission matches the eclipse observations reasonably well in the closed-field streamer regions, although it slightly underestimates the \ion[Fe x] brightness there. The model likewise underestimates the brightness of \ion[Fe xi] and \ion[Fe xiv] in the polar coronal holes. In \cite{Boe2021a}, the K corona of the same PSI MHD model was compared to the eclipse data, and was found to be an excellent match -- implying that the differences between the eclipse and model line emission are not due to a density effect. Instead, these differences are most likely due to a slightly too low energization of the coronal plasma in the model, especially in the polar coronal holes. There could also be a freeze-in/non-equilibrium effect which is shaping the ion distributions in a non-trivial manner (see Section \ref{modelTe}). We intend to explore the exact relationship of the inferred temperatures to the underlying plasma state and the implied systematics in another paper, as it is beyond the scope of the work presented here.

\par

Another important finding of this work, is that \ion[Fe xi] emission is clearly detectable out to at least 3.4 \Rs \ with relatively small telescopes and only a couple of minutes of observing time -- at an eclipse that was near solar minimum where the coronal density (and thus brightness) is at its lowest. It is therefore likely that one could observe line emission to an even higher helioprojective distance at eclipses. We intend to deploy instruments with wider field of views in the future to expand on the maximum distance that line emission can be measured. Further, this finding suggests that space-based coronagraphs which have rather large occulters ($\approx$ 2 \Rs) could be used to measure line emission. Any stray light contamination would be accounted for by subtracting the continuum with an Off-band filter as well (see Section \ref{Eclipse}), since the stray light would be effectively the same over a small wavelength shift between the On- and Off-band filters.

\par
This work also strengthens the value of the forthcoming line emission observations from the ground-based UCoMP instrument \citep{Tomczyk2021}, which can perform similar analysis to that done here, but with a much larger time baseline. While the ground-based observations cannot probe to as high of a helioprojective distance due to the sky brightness of the Earth, they could still explore the line emission and $T_e$ time variations in the corona -- as \cite{Boe2020a} demonstrated for multiple observing sites spaced out over the United States during the 2017 TSE with similar instrumentation used in this work. The calibration and line width corrections made here (see Section \ref{LineWidth}) can offer a means to measure the absolute brightness (and $T_e$) of additional lines with UCoMP in the near future.

\par
Finally, as planned manned lunar missions ramp up in the next decade or so, it may be possible to use the Earth as an occulter during lunar eclipses (i.e. a total solar eclipse on the Moon), or using the Moon as an occulter while in lunar orbit as proposed by \cite{Habbal2013}. The extent of line emission here implies that despite the large size of the Earth, such data would be useful for achieving long exposures of line emission in the corona. As the Earth moves from one side of the corona to the other, the observations could focus on the opposite side of the corona for half of the eclipse duration (which is hours on the Moon, compared to minutes on the Earth). Rather small payloads, similar to the ones used here, could be deployed to the lunar surface or to orbiters such as the planned Lunar Gateway station to take advantage of the multiple lunar eclipses which occur every year.

\subsection*{Acknowledgments}
We thank Judd Johnson and Pavel \v Starha, who acquired the narrowband data at the Rodeo, Argentina observing site during the 2019 July 2 total solar eclipse. 
\par
The K-Cor data were courtesy of the Mauna Loa Solar Observatory, operated by the High Altitude Observatory, as part of the National Center for Atmospheric Research (NCAR). NCAR is supported by the National Science Foundation.
\par
Observables presented in this paper and other eclipse data from our group can be found at: \url{https://www.ifa.hawaii.edu/SolarEclipseData/}. The PSI MHD model eclipse predictions can be found here: \url{https://www.predsci.com/corona/} 
\par
Financial support was provided to B. B. by the National Science Foundation under Award No. 2028173. S. H. and the 2019 eclipse expedition were supported under NASA grant NNX17AH69G and NSF grant AST-1733542 to the Institute for Astronomy of the University of Hawaii. C. D. was supported by the NASA Heliophysics Supporting Research and Living With a Star programs (grants 80NSSC18K1129 and 80NSSC20K0192). M. D. was supported by the Grant Agency of Brno University of Technology, project No. FSI-S-20-6187.

\bibliographystyle{apj}
\bibliography{2019LineEmission_R1_ArXiv}
\end{document}